# Observation of polarization domain wall solitons in weakly birefringent cavity fiber lasers

H. Zhang, D. Y. Tang *, L. M. Zhao, X. Wu

*School of Electrical and Electronic Engineering, Nanyang Technological University, Singapore 639798*

*Corresponding author: edytang@ntu.edu.sg

We report on the experimental observation of two types of phase-locked vector soliton in weakly birefringent cavity erbium-doped fiber lasers. While a phase-locked dark-dark vector soliton was only observed in fiber lasers of positive dispersion, a phase-locked dark-bright vector soliton was obtained in fiber lasers of either positive or negative dispersion. Numerical simulations confirmed the experimental observations, and further showed that the observed vector solitons are the two types of phase-locked polarization domain-wall solitons theoretically predicted.

Soliton formation is a fascinating phenomenon that has been observed in many diverse physical systems [1]. In the context of optics, soliton formation in the single mode fibers (SMFs) has attracted the most attention. Light propagation in a SMF is governed by the nonlinear Schrödinger equation (NLSE), which admits formation of either bright or dark optical solitons depending on the sign of the fiber dispersion. When the vector nature of light is considered, light propagation in a SMF is then governed by the coupled NLSEs, which have much richer dynamics than the scalar NLSE. It has been theoretically shown that polarization coupling of light in a SMF could lead to formation of various types of vector solitons, including the bright-bright [2, 3], dark-dark [4], and dark-bright [5] types of vector solitons. Formation of the incoherently coupled bright-bright and dark-dark types of vector solitons in birefringent SMFs has already been experimentally demonstrated [6, 7].

A fiber laser has its cavity mainly made of SMFs. Although light propagation in a fiber laser cavity is also subject to actions of other cavity components, such as the laser gain and cavity output coupler, it was found that optical solitons could still be formed in a fiber laser, and as far as the spectral bandwidth of the formed solitons is far narrower than the gain bandwidth, the dynamics of the formed solitons could be well described by the NLSE, or the coupled NLSEs when the cavity birefringence needs to be considered [8, 9]. A fiber laser provides a unique nearly conservative system for the optical soliton studies.

Although majority of the theoretically predicted optical solitons in SMFs have been experimentally confirmed, to the best of our knowledge, no experimental observation on the phase-locked polarization domain-wall solitons (PDWSs) has been reported.

Formation of phase-locked PDWSs in diffractive or dispersive Kerr media was first theoretically predicted by Haelterman and Sheppard [10]. They showed that Kerr media could sustain localized structures separating domains of different nonlinear polarization eigenstates. Two types of phase-locked PDWSs, one in the form of elliptically polarized dark-dark vector solitons, and the other one in the form of dark-bright vector solitons exist in fibers of normal dispersion. Although not named as PDWS, Christodoulides had also theoretically predicted a type of bright-dark vector soliton in weakly birefringent fibers [11]. He had given the pulse profiles of the dark and bright solitons, and pointed out that the solitons have the same pulse width, which is uniquely determined by the fiber group velocity dispersion (GVD) and strength of fiber birefringence. Based on our experimental studies, we suspect that the dark-bright vector soliton predicted by Christodoulides could be a special case of the dark-bright domain-wall solitons predicted by Haelterman and Sheppard. S. Pitois et al have experimentally investigated polarization modulation instability (PMI) in a birefringent fiber of normal dispersion [12]. Although in their experiment a kind of PMI induced fast modulation structure was observed, no PDWSs were identified. The difficulties in observing the PDWSs in a SMF are that the fiber birefringence must be sufficiently small and maintained constant over a long distance, furthermore, coupling between the two polarization components must be strong enough.

In a previous experiment we have successfully demonstrated a high-order polarization locked bright-bright vector soliton in a mode-locked fiber laser [13]. To obtain the polarization locked bright-bright vector soliton the averaged cavity birefringence must also be kept sufficiently small. We note that previous experimental studies on the polarization dynamics of fiber lasers have shown that cavity feedback could significantly enhance the polarization coupling of light [14]. Worth of mentioning is that Williams and Roy had demonstrated a kind of anti-phase square-wave fast polarization dynamics in an erbium-doped fiber ring laser [15]. The antiphase square-wave pulses observed by them are reminiscent of the polarization domain-wall structures. It is a well-known fact that a mode locked pulse in an anomalous dispersion fiber laser can be automatically shaped into a NLSE soliton. Considering the nature of PDWS formation in SMFs, it is to expect that under suitable conditions the antiphase square-wave pulses could be further shaped into PDWSs.

We designed two fiber lasers to investigate the formation of PDWSs. The overall laser setup is schematically shown in Fig. 1. The laser cavity is a simple fiber ring comprising a segment of erbium-doped fiber (EDF) as the laser gain medium, a wavelength-division-multiplexer for coupling in the pump light, and a 10% fiber output coupler for the laser output. A fiber-pigtailed polarization independent isolator and an in-line polarization controller are inserted in the cavity to force the unidirectional operation of the ring and to fine-tune the linear cavity birefringence. To study the effects of cavity dispersion on the PDWSs, we have made one laser with the purely anomalous dispersion fibers while the other the purely normal dispersion fibers. The laser emission along the two orthogonal principal polarization directions of the cavity is resolved using a fiber polarization beam splitter (PBS).

The negative dispersion fiber laser has a cavity length of 14.7m, with 6.4 m EDF of GVD of 10 (ps/nm)/km and 8.3 m SMF of GVD of 18 (ps/nm)/km. When pump power of the laser was increased to about 4~5 times of the laser threshold, the laser emission exhibited a periodic polarization switching between the two orthogonal principal polarization directions. Under the polarization resolved measurement it is represented as an antiphase square pulse emission, which was explained previously as caused by the delayed cavity feedback and the polarization mode competition [15]. Experimentally we found that associated with the polarization switching of the laser emission, a PDWS was also formed, indicating that the laser emission along both of the two principal polarization directions were stable. At fixed pump strength, tuning the paddle orientation of the polarization controller, which corresponds to changing the average linear cavity birefringence, the square pulse width could be continuously changed. Multiple square-pulses could also be obtained under strong pumping. However, due to the intrinsic scalar modulation instability caused by the negative dispersion fibers, too strong pumping also resulted in formation of the NLSE type solitons within in the square pulses, which destabilizes the square pulses. Therefore, we have restricted us to the cases where no NLSE solitons appeared. Once the periodic square pulse emission of the laser was obtained, we then carefully decreased the linear cavity birefringence, which was experimentally detected by monitoring the lasing wavelength difference along the two polarization directions. In this way we could continuously decrease the square-pulse width, eventually a stable state as shown in Fig. 2 could be obtained. On the oscilloscope traces (Fig. 2a) it is to see that while along one principal polarization direction the laser emitted a train of pulses (lower trace of Fig. 2a), along the orthogonal polarization direction the laser emitted strong CW, and on the CW background there is a train of intensity dips (upper trace of Fig. 2a). Each intensity dip corresponds temporally to an intensity pulse, and the dark-bright pulses repeat with the fundamental cavity repetition rate. Fig. 2b shows a zoom-in of the bright and dark pulses. They have almost the same pulse profiles and widths that are limited by the resolution of the electronic detection system. Using a commercial autocorrelator we also measured the autocorrelation trace of the bright pulses. A typical result is shown in the inset of Fig. 2c. The autocorrelation trace has a good $Sech^2$-profile with a FWHM of about 4.8 ps, which gives that the pulse width is 3.1ps. Due to the low pulse repetition rate, the pulse width of the dark pulses could not be measured with the autocorrelator.

Fig. 2c shows the polarization resolved optical spectra of the laser emission. The soliton feature of the pulses is characterized by the appearance of the clear spectral sidebands on their spectra. We note that even the dark pulses have spectral sidebands, and especially whose spectral sidebands have the same locations as those of the bright pulses. The 3dB spectral bandwidth of the bright pulses is ~ 0.9 nm. Therefore, the bright solitons are near transform-limited. There is no obvious spectral difference between the dark and bright pulses, furthermore, no polarization evolution frequency [8] was observed on neither the bright nor the dark pulses, which clearly suggests that the phases of the pulses are locked. Therefore, the dark-bright pulses shown in Fig. 2a constitute a phase-locked dark-bright vector soliton (DBVS).

Experimentally the DBVS could be obtained at various pump strengths. The higher the pump power, the easier is the state obtained. Under strong pumping occasionally multiple DBVSs have also been obtained. However, different from the cases of multiple bright vector solitons formed in a fiber laser, the multiple DBVSs could have different soliton energies, represented by their different pulse heights in the same oscilloscope trace. Energy of the bright and dark pulses in a vector soliton is always correlated.

Corresponding to a bright soliton with a weak pulse, the dark soliton also has a shallow dip. In the case that the formed DBVSs have different soliton energies, slow relative movement between them was observed. Two vector solitons were observed to collide and then merge together.

The positive dispersion fiber laser has a cavity length of 9.2 m. It comprises 5 m EDF of GVD of -32 (ps/nm)/km and 4.2m DCF of GVD of -4 (ps/nm)/km. Periodic antiphase square-wave pulses have also been observed in the laser. However, because no intrinsic modulation instability exists, no fine structures were observed within the square pulses even under strong pumping. Exactly like those square-wave pulses observed in the negative dispersion fiber laser, decreasing the linear cavity birefringence the width of the square pulses could be reduced. With an appropriate selection of the cavity parameters, fundamental DBVSs have also been revealed in the laser, which confirms Christodoulides's prediction that formation of the DBVSs is independent of the fiber dispersion [11]. Apart from the DBVSs, in the laser in a narrow cavity birefringence regime we have further obtained a dark-dark type of vector soliton as shown in Fig. 3. In the state the dark solitons were found static in the CW background, which is clearly different from the DBVSs observed. Based on the polarization resolved soliton spectra and the evolution of the dark solitons with respect to the cavity roundtrips, we further confirmed that the phases of the dark solitons were locked. They constitute a fundamental phase-locked dark-dark vector soliton.

To confirm the dark-bright and dark-dark vector soliton formation in the fiber lasers, we further numerically simulated the operation of the lasers. To faithfully simulate the nonlinear light propagation in the weakly birefringent fibers and the laser cavity feedback, we used a model as described in [16]. Briefly, we circulated the light within a simulation window in the laser cavity. The light propagation in the cavity fibers was described by the coupled extended NLSEs. When the light propagated in the gain fiber, we also considered the gain effects. We had always started the calculations with an arbitrary weak pulse. After one cavity roundtrip of calculation, the result was then used as input for the next roundtrip of calculation, until a stead state of the light field was reached. To make the simulations possibly close to the experimental situations, we used the actual laser cavity lengths and the following parameters for the fibers: $\gamma=3$ $W^{-1}km^{-1}$, $k'''= -0.13$ $ps^3/km$, $k''_{SMF}= -23$ $ps^2/km$, $k''_{EDF1}= -13$ $ps^2/km$, $k''_{DCF}= 5.2$ $ps^2/km$, $k''_{EDF2}= 41.6$ $ps^2/km$. For the gain fibers we used the gain bandwidth $\Omega_g = 16$ nm, and gain saturation energy $P_{sat,EDF1} = 50$ pJ, $P_{sat,EDF2}= 500$ pJ.

Independent of the sign of cavity dispersion, stable DBVSs could indeed be reproduced in our simulations. Fig. 4a and Fig. 4b shows one of the calculated DBVSs, obtained under a cavity linear birefringence of $\Delta n = n_u - n_v = 5.3\times10^{-9}$. Both solitons co-propagate in the cavity, and their phases are locked along the propagation. Numerically it was noticed that in order to obtain a stable phase-locked DBVS, the cavity birefringence must be sufficiently small. Moreover, the pump power must be in an appropriate range. Spectral sidebands can also be identified on the calculated soliton spectra. Like the experimental observations, the sidebands of both solitons have the same locations on the soliton spectrum. Stable dark-dark vector solitons were also numerically obtained in the positive dispersion fiber laser, as shown in Fig. 4c. Numerically it was confirmed that the dark-dark vector solitons are phase locked. We note that the dips of all the numerically calculated dark solitons have reached zero, which indicates that they are black solitons.

However, the calculated bright soliton has a non-zero background. These features of the numerically calculated vector solitons well match those of the PDWSs theoretically predicted in [10], suggesting that the observed vector solitons are formed due to the coherent cross coupling between the two polarization components.

Finally, we note that the spectral bandwidth of both the experimentally observed and the numerically calculated vector solitons are far narrower than the gain bandwidth. We had also experimentally observed formation of the NLSE solitons in the negative dispersion fiber laser under strong pumping. The formed NLSE solitons are embedded in the square-wave pulses. Spectral bandwidth of the NLSE solitons was found much broader than those of the domain wall solitons. In addition, PMI was occasionally observed in our experiments. Its appearance had a higher threshold than that of the PDWSs. PMI introduced a high frequency intensity modulation on the square pulses of the lasers. Because of the much narrower spectral bandwidth of the observed PDWSs than the laser gain bandwidth, the effect of the laser gain is purely to balance the cavity loss.

In conclusion, we have first experimentally observed a fundamental phase-locked dark-bright and dark-dark vector soliton in weakly birefringent cavity fiber lasers, respectively. The vector solitons were formed due to the strong cross polarization coupling of light in the fiber lasers. They constitute the two types of the fundamental PDWS theoretical predicted for the coupled NLSEs. Since the coupled NLSEs describe a wide range of physical systems, such as the Bose-Einstein condensate, we believe our experimental observation could be of fundamental as well practical importance.

Figure captions:

Fig.1. (Color online) Setup of the fiber lasers. WDM: wavelength division multiplexer. EDF: erbium doped fiber. PC: Polarization controller. PBS: Polarization beam splitter.

Fig. 2. (Color online) A typical dark-bright vector soliton emission of the lasers. (a) Oscilloscope traces. (b) Zoom-in of the dark-bright pulses. (c) Polarization resolved optical spectra; inset: Autocorrelation trace of the bright pulses.

Fig. 3. (Color online) Oscilloscope traces of a typical dark-dark vector soliton emission of the positive dispersion fiber laser.

Fig. 4. (Color online) The dark-bright and dark-dark vector soliton states numerically calculated. (a) Evolution of the dark-bright vector soliton with the cavity roundtrips. (b) Optical spectra of the dark and bright solitons. Inset: soliton profiles of the dark and bright solitons. *Gain=120*. (c) Soliton profiles of the dark solitons numerically calculated. *Gain=200, $\Delta n = n_u - n_v = 1.7 \times 10^{-9}$*.

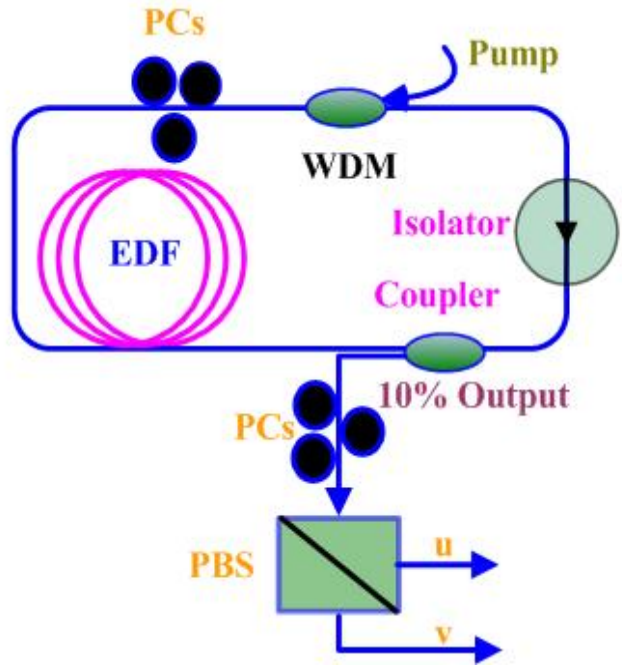

Fig. 1    H. Zhang et al.

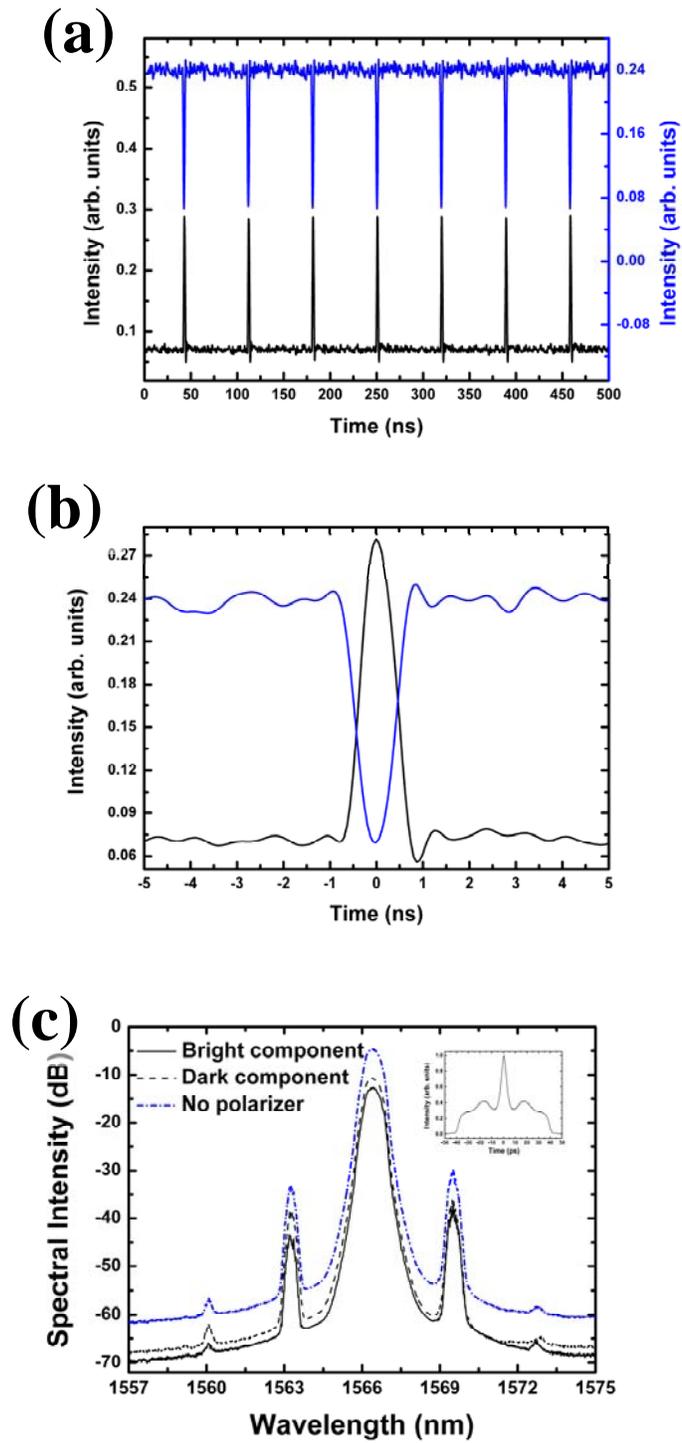

Fig. 2  H. Zhang et al.

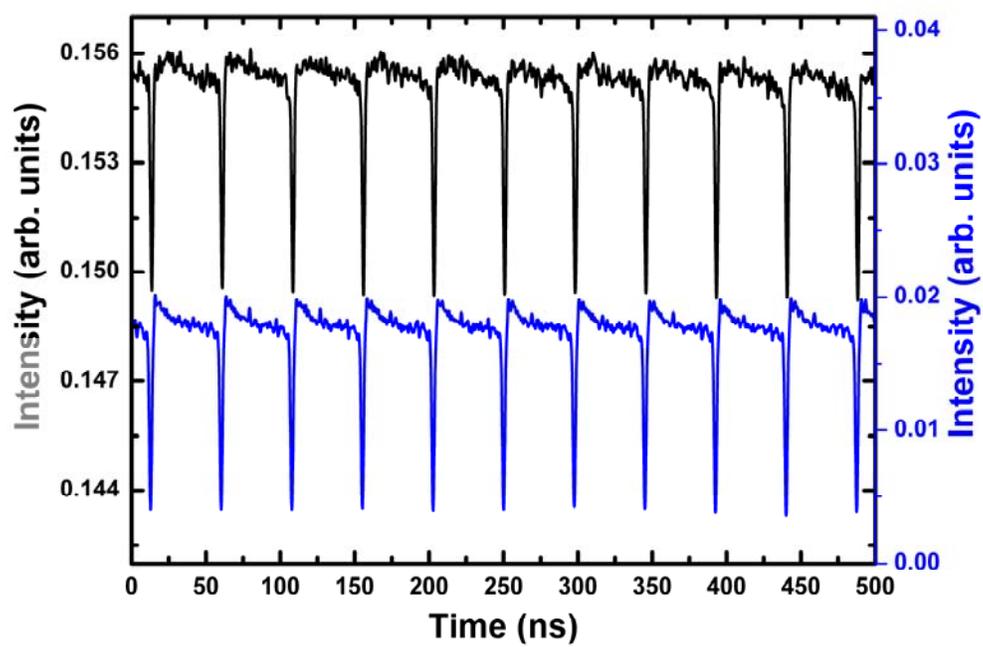

Fig. 3　H. Zhang et al.

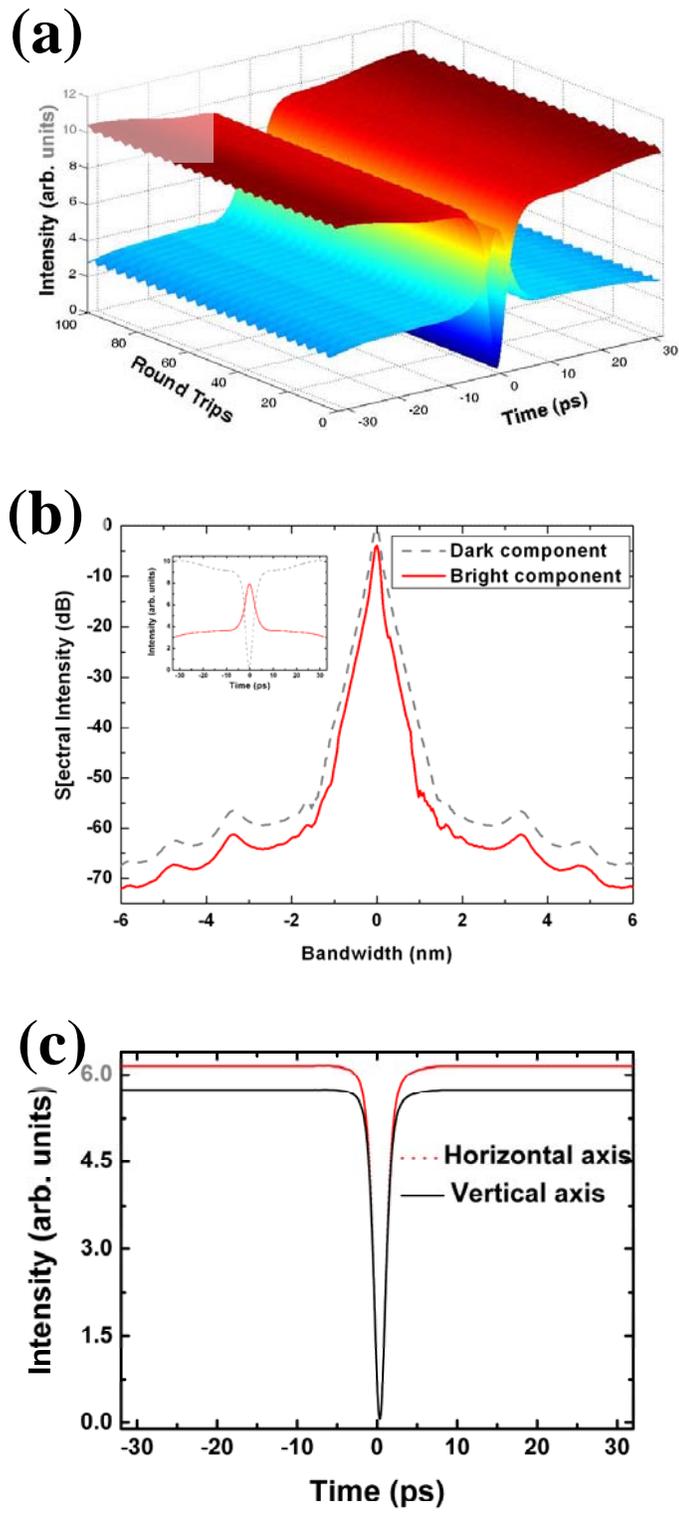

Fig. 4　H. Zhang et al.

16